\documentstyle[prl,aps,multicol,epsf]{revtex}

\begin{document}
\draft
\title{Quantizing Charged Magnetic Domain Walls: Strings on a Lattice}
\author{H. Eskes, R. Grimberg, W. van Saarloos and J. Zaanen}
\address{Institute Lorentz, Leiden University, P.O. Box 9506,
         NL-2300 RA Leiden, The Netherlands}
\date{\today}
\maketitle

\begin{abstract}
The discovery by Tranquada {\em et al.} of an ordered phase of charged domain
walls in the high-$T_c$ cuprates leads us to consider the possible existence
of a quantum domain-wall liquid. We propose minimal models for the
quantization, by meandering fluctuations, of isolated charged domain walls.
These correspond to lattice string models. The simplest model of this kind,
a directed lattice string, can be mapped onto a quantum spin chain or on a
classical two-dimensional solid-on-solid surface model. The model exhibits
a rich phase diagram, containing several rough phases with low-lying
excitations as well as ordered phases which are gapped.
\end{abstract}


\begin{multicols}{2}

The study of high-$T_c$ superconductors has caused a crisis of the
conventional paradigm of metal physics, the Fermi-liquid
theory which insists that the current in metals is carried by quasiparticles
with all properties of electrons except that their effective interactions
vanish at the Fermi surface. It seems now widely agreed
that this theory fails fundamentally in the context of the high-$T_c$
cuprates.
Indeed, a  strong case can be made that any theory starting with
a particle-like vacuum is bound to fail \cite{emery}. From
a theoretical viewpoint this suggests that the Fermi-liquid fixed point can be
unstable and in recent years there have been many investigations aimed at
the possible breakdown of Fermi-liquid theory or at identifying new phases of
strongly correlated fermions.

The present study of quantum domain walls
is motivated by recent compelling experimental evidence \cite{tranquada}
suggesting that nature has indeed chosen to realize a collective phase whose
basic ingredient is an intuitively simple and easily identifiable
many-particle bound state: a charged domain wall.
Microscopically these domain walls in two dimensions consist of
holes bound in a linear string-like fashion, separating antiferromagnetically
ordered regions. Across a wall the antiferromagnetic order parameter points in
opposite directions  (see Fig. 1). Tranquada {\it et al.} find
a freezing of these domain walls in a so-called striped phase at a doping
concentration of $x=1/8$, in the middle of the superconducting regime
in $La_{1.48}Nd_{0.4}Sr_{0.12}CuO_4$ \cite{tranquada}.

These domain walls were found some time ago to be the {\em generic}
semi-classical (spin $S \rightarrow \infty$)
mean-field solutions
\begin{figure}
\epsfxsize=.8\hsize
\vspace{0.5ex}
\hspace{2.5ex}\epsffile{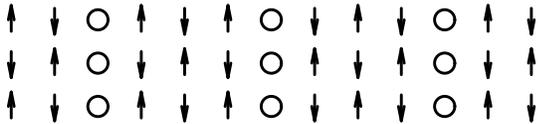}
\vspace{1.5ex}
\narrowtext
\caption[]{
Schematic illustration of a domain wall of holes separating two
antiferromagnetically ordered domains. Note that in reality the hole density
may vary more gradually across the wall.}
\end{figure}
\noindent of models for doped Mott-Hubbard
insulators \cite{zagun}.
As is well known, the motion of a hole in an antiferromagnetic background is
hampered by the spin order. In domain walls the holes still have room to
fluctuate, so that their kinetic energy is low, while at the same time the
number of broken (spin) bonds per hole is small. Domain walls also survive
in more sophisticated treatments\cite{laterwork}, and their formation is
favored by Coulomb\cite{emery2} and electron-phonon interactions\cite{zali}.
Domain wall striped phases
were found before in two dimensional, doped $La_2NiO_4$\cite{tranquada1},
where one is most likely approaching the classical limit more closely
\cite{zali} than in the cuprates.

As argued in the paper by Tranquada {\it et al.}\cite{tranquada}, the
occurrence of the striped phase in the cuprate is suggestive of dominating
domain wall correlations in the fluid (metallic, superconducting) phases as
well. After all, the freezing is explained \cite{tranquada}
in terms of a simple commensuration
effect, which is inactive in the absence of charged domain wall correlations.
Moreover, the spin fluctuations in the metallic state \cite{aeppli}
closely match the static spin incommensurations in the ordered state
\cite{barzykin}.

These observations force one to consider the possibility that the metallic
state is  a fluid of charged domain walls: a phase with strong dynamical hole
correlations which are reminiscent of a disordered domain wall network.
We argued before \cite{zahosa} that if the (collective) dynamics of such a
domain wall fluid is assumed to be driven by thermal fluctuations, one
obtains a natural explanation for the anomalous magnetic dynamics in the
metallic state. This analysis indicates that at not too low temperatures
there should be a cross-over to a low temperature regime where the fluid
becomes dominated by quantum fluctuations --- obviously, because the ground
state is a superconductor and not a striped phase.
Virtually all questions  concerning the precise nature
of this conjectured {\em quantum domain wall fluid} are, at present, wide open.

In this paper, we focus on the quantum meandering fluctuations of a single
wall on a lattice. In developing a model, we are guided by several
observations. ({\em i}\ )
{}From Hartree-Fock calculations \cite{zagun}, it is known that the energy
associated with adding or removing a hole from a domain wall is rather high.
Hence, we consider the number of holes in a wall to be fixed. ({\em ii}\ )
The well-known fact that long wavelength shape fluctuations of fluid
interfaces (i.e. capillary waves) and membranes have a low energy,
suggests that these are important low energy modes of a domain wall as well.
Moreover, mean field calculations indicate that the ``stiffness'' associated
with these modes is small:
kinks have a low creation energy, while their tunneling rate to neighboring
sites is appreciable \cite{vierrice}. ({\em iii}\ ) From the theory of
interfaces, it  is well known that meandering fluctuations are not properly
incorporated in a mean field theory, and in order to understand the
commensurability effect of a lattice (roughening transition!), one has to
start with a proper microscopic lattice model.
({\em iv}\ ) Quantum mechanical domain wall fluctuations reduce the
kinetic energy of a hole. For walls in an antiferromagnetic background
(Fig.~1), the low energy fluctuations without spin frustration are those
where the holes preserve their string-like order (as beads on a string),
in other words, where the wall does not break up into two disconnected pieces.
({\em v}\ ) The latter fluctuations dominate the dynamics of spins
near a wall. It is therefore reasonable to neglect in a first stage the
interaction between spin waves in the
antiferromagnetic domains and the wall degrees of freedom.

The model we propose for the strong hole-binding limit of quantum domain walls
is that of {\em quantum lattice strings}. These are connected strings of
``holes'' on a square lattice, successive holes $l$ and $l+1$ having
either a distance 1 or $\sqrt{2}$. Let $(\eta^x_l,\eta^y_l)$ be the position
of hole $l$. We write the classical interaction as a sum of nearest and
next-nearest neighbor discretized string-tensions,
\begin{eqnarray}
{\cal H}^{Cl} &=& \sum_l \left[ K \delta( | \eta^x_{l+1} - \eta^x_l | - 1 )
   \delta( | \eta^y_{l+1} - \eta^y_l | - 1 )   \right. \nonumber \\
&+&  \sum\limits_{i,j = 0}^{2} \left. L_{ij}
\delta( | \eta^x_{l+2} - \eta^x_{l}| - i)
\delta( | \eta^y_{l+2} - \eta^y_l | - j ) \right]    \nonumber \\
&+&  M \sum\limits_{l,m} \delta( \eta^x_{l} - \eta^x_{m} )
                       \delta( \eta^y_{l} - \eta^y_{m} ) .
\end{eqnarray}
The relevant local configurations are shown in Fig.~2a. The last term is an
excluded volume type interaction; the physically relevant limit is
$M \rightarrow \infty$, so that holes cannot occupy the same site
($L_{00} $ is now irrelevant). We neglect extreme
curvature, $L_{10}=\infty$, and choose $L_{20}=0$. Therefore the model
is parameterized by $K,L_{11},L_{21}$, and $L_{22}$. The strings are
quantized by introducing conjugate momenta $\pi^{\alpha}_l$,
$[ \eta^{\alpha}_l, \pi^{\beta}_{m} ] = i \delta_{l,m} \delta_{\alpha,\beta}$.
A term $e^{ i n \pi^{x}_l}$ will cause hole $l$ to hop
a distance $n$ in the $x$-direction. Therefore the hopping contribution
in its simplest, nearest-neighbor form \cite{twohops} is,
\begin{equation}
{\cal H}^{Qu} = 2t \sum\limits_{l,\alpha} P^\alpha_{Str}(l) \cos(\pi^\alpha_l).
\end{equation}
\begin{figure}
\epsfxsize=.8\hsize
\vspace{0.5ex}
\hspace{2.5ex}\epsffile{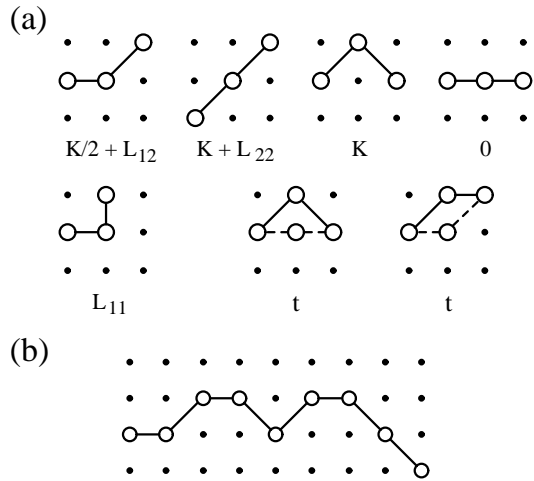}
\vspace{1.5ex}
\narrowtext
\caption[]{(a) Energies (of the central hole) and tunneling amplitudes of
the various local-bond configurations. The tunneling process is between the
dashed configuration and the one drawn in full.
(b) Typical configuration of a rough directed string on a square lattice.}
\end{figure}
Here $P^\alpha_{Str}(l)$ is a projector restricting the motion of hole $l$ in
the $\alpha$-direction to string configurations (Fig.~2a) \cite{reconnect}.
By means of the Suzuki-Trotter mapping the quantum problem ${\cal H}^{Cl} +
{\cal H}^{Qu}$ translates to the problem of two coupled 1+1 D restricted
solid-on-solid (RSOS) models. The two RSOS height flavors correspond to
$\eta^x_{l,k}$ and $\eta^y_{l,k}$, $k$ being the imaginary-time Trotter index.
The ``steps'' $\eta^\alpha_{l+1,k}-\eta^\alpha_{l,k}$ and
$\eta^\alpha_{l,k+1}-\eta^\alpha_{l,k}$ are restricted to $0,\pm 1$.

An important simplification occurs by restricting the allowed configurations
to those with $\eta^x_l = l$. In this {\it directed}
string problem the bonds between neighboring holes
always step to the right (see Fig.~2b) --- in interface language: the
strings have no ``overhangs''. By using the Suzuki-Trotter or transfer
matrix formulation, one can write the ground state problem of a directed
lattice string in terms of the  statistical mechanics of a classical
solid-on-solid surface model \cite{dennijs}. In this mapping, one may think
of the quantum lattice string as tracing out a two-dimensional world sheet in
space-time; this formulation has the advantage that it allows a rather natural
and direct translation of various phases of the two-dimensional surface
problem into those of quantum strings --- e.g. rough surfaces correspond
to lattice strings whose quantum fluctuations are so strong that they
meander, even in the ground state.

The equivalence between RSOS problems and spin algebras was extensively
discussed in the seminal work by den Nijs and Rommelse \cite{dennijs}.
As the links between neighboring holes can only point in three directions
(Fig.~2a), the Hamiltonian can equivalently be formulated in terms of
a spin-1 model, with $S_l^z = 1,0,-1$ corresponding with the link between
site $l$ and $l+1$ pointing up, horizontal and down, respectively.
In terms of the string-tension parameters,
\begin{eqnarray}
{\cal H} & = & \sum_l \left[  ( { { L_{22}} \over 2} -
2 L_{21} ) ( S^z_l S^z_{l-1} )^2
+ { {L_{22}} \over 2 } S^z_l S^z_{l-1} \right. \nonumber \\
& &   \left. + (  K + 2L_{21} ) ( S^z_l)^2 + { t \over 2} (S^+_l S^-_{l-1}
+ S^-_{l} S^+_{l-1}) \right].
\end{eqnarray}
In the special case $L_{22}=4L_{21}$ the Hamiltonian reduces to the
well-known spin-1 chain with single-site anisotropy $D=(K+2L_{21})/t$ and
Ising anisotropy $J=L_{22}/2t$. The phase diagram as given by
den Nijs and Rommelse \cite{dennijs} is shown in the inset in Fig.~3.
However, for $L_{22} > 4L_{21}$ we find several extra phases to occur. Using a
combination of quantum-Monte Carlo, exact diagonalization and knowledge of the
critical behavior at the various transitions
\cite{dennijs,eskes} we arrive at the zero-temperature phase diagram for
$L_{22}/2-2L_{21}=5$ shown in Fig.~3.

There are several ordered phases in Fig.~3. They are listed in Table 1.
For large negative $L_{22}$ diagonal
walls are found (phase I) and for large $K$ the walls are horizontal and flat
(phase II). Positive $L_{22}$ (phase III) favors (not very realistic) zigzag
walls (or antiferromagnetism in spin language).
Apart from these, two new phases occur. Phase VIII corresponds to a
$22.5^\circ$ slanted phase of alternating diagonal and horizontal bonds
(alternating $S^z=1$ and $S^z=0$). Phase X is similar
but now the diagonals themselves have an
alternating up step, down
\begin{figure}
\epsfxsize=1.05\hsize
\vspace{0.5ex}
\hspace{-6ex}\epsffile{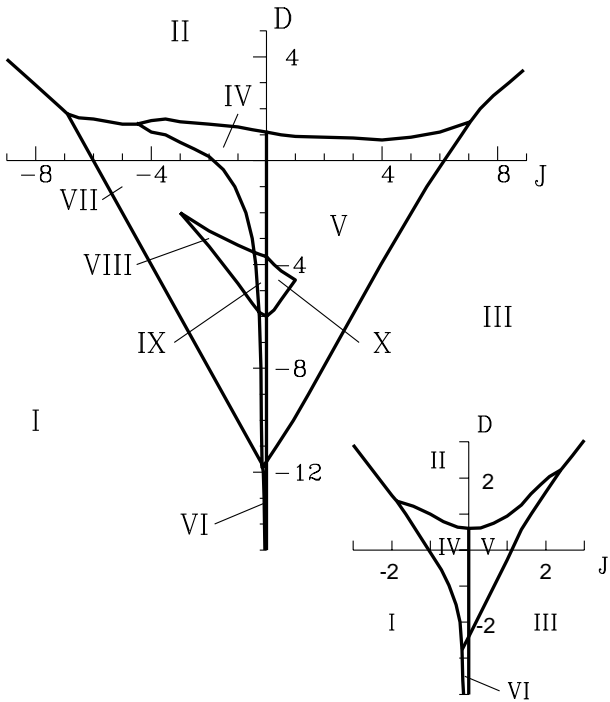}
\vspace{1.5ex}
\narrowtext
\caption[]{Phase diagram of the directed lattice-string problem with
$L_{22}/2-2L_{21}=5$, as a function of $J=L_{22}/2$ and $D=K+2L_{21}$.
The inset shows the phase diagram for $L_{22}=4L_{21}$.}
\end{figure}
\noindent step order. The excitation spectra of these phases are gapped.
Alternating order of diagonal and horizontal bonds is favored by
$L_{22}/2 - 2L_{21} > 0$.

The remaining phases are quantum disordered. They can be characterized by the
presence and/or absence of order of the diagonal and horizontal bonds.
Starting from the flat phase (phase II) and lowering $K$ ($L_{22}<0$)
the quantum meandering fluctuations become dominant and the lattice
string undergoes a roughening transition \cite{fisherweeks}.
Like rough two-dimensional interfaces, the
vertical  displacements $\eta^y_l$ of the hole have logarithmically
diverging spatial correlations
$\langle (\eta^y_l-\eta^y_m)^2\rangle \sim \ln |l-m|$.
These rough strings have low lying excitations (capillary waves in interface
language, spin waves in spin language). In phase IV both the diagonal
and horizontal bonds are disordered. Phase VI occurs at large negative $K$.
Here only virtual pairs of horizontal bonds can occur.
The model can be mapped on the anisotropic
Heisenberg spin-1/2 XXZ model and phase VI corresponds to an anisotropy
parameter $-1 < \Delta < 1$ of this model. Phase V is the disordered flat
or Haldane phase.
In this region up steps of the strings are followed by down steps,
but the location of these up and down steps has no long range order because the
horizontal bonds are not ordered. As a result, the lattice strings
are macroscopically flat. In terms of the height variables the order
parameter is local and is given by
$\langle \exp(i\pi\eta^y_l) (\eta^y_{l+1}-\eta^y_l) \rangle$ \cite{dennijs}.

\vspace{0.3cm}
\setlength{\unitlength}{0.1cm}
\begin{table}
\narrowtext
%\hspace{0.5cm}
\caption{ A schematic representation of the different phases.  }
\vspace{0.2cm}
%\hspace{0.5cm}
\begin{tabular}[t]{ccc}
     Phase         &  String  &  Spin 1   \\
\hline
       I           &
\begin{picture}(10,10)
 \multiput(0,0)(2,2){4}{\circle*{0.9}}
\end{picture}
                              & ++++++++   \\
       II          &
\begin{picture}(18,4)
 \multiput(0,0)(2,0){9}{\circle*{0.9}}
\end{picture}
                              &  0 0 0 0 0 0 0 0    \\
       III         &
\begin{picture}(18,6)
 \multiput(0,0)(4,0){5}{\circle*{0.9}}
 \multiput(2,2)(4,0){4}{\circle*{0.9}}
\end{picture}
                              & +$-$+$-$+$-$+$-$     \\
       IV          &
\begin{picture}(20,8)
 \put(0,0){\circle*{0.9}}
 \put(2,2){\circle*{0.9}}
 \put(4,2){\circle*{0.9}}
 \put(6,0){\circle*{0.9}}
 \put(8,2){\circle*{0.9}}
 \put(10,2){\circle*{0.9}}
 \put(12,4){\circle*{0.9}}
 \put(14,4){\circle*{0.9}}
 \put(16,2){\circle*{0.9}}
 \put(18,4){\circle*{0.9}}
\end{picture}
                              & +0$-$+0+0$-$+   \\
       V           &
\begin{picture}(16,6)
 \put(0,2){\circle*{0.9}}
 \put(2,0){\circle*{0.9}}
 \put(4,2){\circle*{0.9}}
 \put(6,2){\circle*{0.9}}
 \put(8,0){\circle*{0.9}}
 \put(10,0){\circle*{0.9}}
 \put(12,0){\circle*{0.9}}
 \put(14,2){\circle*{0.9}}
 \put(16,0){\circle*{0.9}}
\end{picture}
                              & $-$+0$-$0 0+$-$     \\
       VI          &
\begin{picture}(16,8)
 \put(0,0){\circle*{0.9}}
 \put(2,2){\circle*{0.9}}
 \put(4,0){\circle*{0.9}}
 \put(6,2){\circle*{0.9}}
 \put(8,4){\circle*{0.9}}
 \put(10,2){\circle*{0.9}}
 \put(12,4){\circle*{0.9}}
 \put(14,2){\circle*{0.9}}
 \put(16,0){\circle*{0.9}}
\end{picture}
                              & +$-$++$-$+$-$$-$     \\
       VII          &
\begin{picture}(10,10)
 \put(0,0){\circle*{0.9}}
 \put(2,2){\circle*{0.9}}
 \put(4,2){\circle*{0.9}}
 \put(6,4){\circle*{0.9}}
 \put(8,6){\circle*{0.9}}
 \put(10,6){\circle*{0.9}}
\end{picture}
                              & 0+0++0+0 0     \\
       VIII          &
\begin{picture}(14,10)
 \put(0,0){\circle*{0.9}}
 \put(2,0){\circle*{0.9}}
 \put(4,2){\circle*{0.9}}
 \put(6,2){\circle*{0.9}}
 \put(8,4){\circle*{0.9}}
 \put(10,4){\circle*{0.9}}
 \put(12,6){\circle*{0.9}}
 \put(14,6){\circle*{0.9}}
\end{picture}
                              & 0+0+0+0+0     \\
       IX          &
\begin{picture}(18,8)
 \put(0,0){\circle*{0.9}}
 \put(2,0){\circle*{0.9}}
 \put(4,2){\circle*{0.9}}
 \put(6,2){\circle*{0.9}}
 \put(8,4){\circle*{0.9}}
 \put(10,4){\circle*{0.9}}
 \put(12,2){\circle*{0.9}}
 \put(14,2){\circle*{0.9}}
 \put(16,4){\circle*{0.9}}
 \put(18,4){\circle*{0.9}}
\end{picture}
                              & 0+0+0$-$0+0     \\
       X           &
\begin{picture}(14,6)
 \put(0,0){\circle*{0.9}}
 \put(2,0){\circle*{0.9}}
 \put(4,2){\circle*{0.9}}
 \put(6,2){\circle*{0.9}}
 \put(8,0){\circle*{0.9}}
 \put(10,0){\circle*{0.9}}
 \put(12,2){\circle*{0.9}}
 \put(14,2){\circle*{0.9}}
 \put(16,0){\circle*{0.9}}
 \put(18,0){\circle*{0.9}}
\end{picture}
                              & 0+0$-$0+0$-$0     \\

\end{tabular}
\label{tabel}
\end{table}
There are two new rough phases, again occuring when
$L_{22}/2 - 2L_{21}$ is sufficiently large and positive.
Region VII is a rough slanted phase: up steps are diluted by
horizontal bonds with positional disorder. The average
angle of the string is
smaller than $45^\circ$ and the deviations from the average are again
logarithmic. Phase IX is a rough phase where the even bonds are horizontal
and the odd bonds are diagonal with up-down disorder.
On average, the string is horizontal.

The results summarized in the phase diagram give a clear answer to whether
and how quantum-domain walls can ``melt''\cite{vierrice}: they can roughen
either via a conventional Kosterlitz-Thouless transition
(from phase II to IV), a
first-order KDP transition (phase I and IV) \cite{dennijs} or a
Pokrovsky-Talapov transition (phase I to VII).
Such rough walls in our view are the
building blocks of the conjectured domain-wall fluid and their
low-lying collective excitations may be responsible for anomalous features
of the metallic state\cite{zahosa}.
The structural deformation of the LTT
phase will suppress the diagonal segments, leading to an effective increase
of the $K$ parameter. Therefore we envision that the formation of the stripe
ordering in the $La$ compounds \cite{tranquada} is related to a transition
from phase IV to phase II. At the transition meandering fluctuations become
gapped. However, one should note that even in the most rough phase
(phase IV) the meandering
fluctuations are only logarithmic at low temperatures, and therefore the
domain walls will be very susceptible to wall-wall interactions and long-range
Coulomb forces, which will promote ordering \cite{coppersmith}.
Reconnection effects \cite{reconnect}, however, will
favor a fluid phase and compete with the above ordering mechanisms.
Other crucial questions that need to be addressed is whether the domain wall
scenario is compatible with the existence of fermion-like excitations,
as seen e.g.\ in photoemission experiments and in studies of the $t$-$J$
model, and how exchange with free holes can be incorporated and affects
the behavior.

HE is supported by the Stichting voor Fundamenteel Onderzoek der Materie
(FOM), which is financially supported by the Nederlandse Organisatie voor
Wetenschappelijk Onderzoek (NWO), and JZ acknowledges support by the Dutch
Royal Academy of Sciences (KNAW).

\end{multicols}


\begin{references}
\bibitem{twohops}
More generally, it is natural to take the hopping amplitudes
of the last two configurations in Fig.~2a different.
\bibitem{reconnect}
An interesting and possibly important extension of (1) is
to include terms describing self-avoidance and/or reconnection processes,
the latter leading to string loops splitting off.
\end{references}
\end{document}